\documentclass[conference]{IEEEtran}
\IEEEoverridecommandlockouts
\newcommand{\myparatight}[1]{\smallskip\noindent{\bf {#1}:}~}
\usepackage{cite}
\usepackage{amsmath,amssymb,amsfonts}
\usepackage{graphicx}
\usepackage{textcomp}
\usepackage{booktabs}
\usepackage{multirow} 
\usepackage{xcolor}
\usepackage{graphicx}

\usepackage{xspace}
\usepackage{bm}
\usepackage{algorithm}
\usepackage{algpseudocode}
\usepackage{float}
\usepackage{dblfloatfix}

\newcommand{\alg}{\text{ProtoGuard-SL}\xspace}

\def\BibTeX{{\rm B\kern-.05em{\sc i\kern-.025em b}\kern-.08em
    T\kern-.1667em\lower.7ex\hbox{E}\kern-.125emX}}
\begin{document}

\title{ProtoGuard-SL: Prototype Consistency Based Backdoor Defense for Vertical Split Learning}


\author{
\IEEEauthorblockN{
Yuhan Shui$^{1,*}$, 
Ruohan Jin$^{1,*}$, 
Zhihao Dou$^{3}$, 
Zhiqiang Gao$^{1,2,\dagger}$
}

\IEEEauthorblockA{$^{1}$Computer Science Department, Wenzhou-Kean University, Wenzhou, China}

\IEEEauthorblockA{$^{2}$International Frontier Interdisciplinary Research Institute (IFIRI), Wenzhou-Kean University, Wenzhou, China}

\IEEEauthorblockA{$^{3}$Case Western Reserve University, Cleveland, OH, USA}

\IEEEauthorblockA{\{1337084,1337268,
zgao\}@wku.edu.cn, \{zxd283\}@case.edu}

\IEEEauthorblockA{$^{*}$Equal contribution \qquad $^{\dagger}$Corresponding author}
}

\maketitle

\begin{abstract}
Vertical split learning (SL) enables collaborative model training across parties holding complementary features without sharing raw data, but recent work has shown that it is highly vulnerable to poisoning-based backdoor attacks operating on intermediate embeddings. By compromising malicious clients, adversaries can inject stealthy triggers that manipulate the server-side model while remaining difficult to detect, and existing defenses provide limited robustness against adaptive attacks.
In this paper, we propose \alg, a server-side defense that improves the robustness of split learning by exploiting \emph{class-conditional representation consistency} in the embedding space. Our approach is motivated by the observation that benign embeddings within the same class exhibit stable semantic alignment, whereas poisoned embeddings inevitably disrupt this structure. \alg adopts a two-stage framework that constructs robust class prototypes and transforms embeddings into a prototype-consistency representation, followed by a class-conditional, distribution-free conformal filtering strategy to identify and remove anomalous embeddings. Extensive experiments are conducted on three datasets, CIFAR-10, SVHN, and Bank Marketing, under three different attack settings demonstrate that our method achieves state-of-the-art performance.
\end{abstract}

\begin{IEEEkeywords}
Split Learning,
Backdoor Attacks,
Robust Embedding Defense
\end{IEEEkeywords}

\section{Introduction}
\label{sec:intro}

As data privacy and regulatory constraints intensify, feature-wise data fragmentation across institutions has become increasingly prevalent, where different parties hold complementary features of the same samples but cannot share raw data. To address this challenge, vertical split learning (SL)~\cite{fu2022blindfl,romanini2021pyvertical,liu2020asymmetrical,thapa2022splitfed,poirot2019split,vepakomma2018split,singh2019detailed} has emerged as an effective collaborative learning paradigm and is often viewed as an important variant of federated learning (FL)~\cite{mcmahan2017communication}, albeit with substantially different training protocols and security assumptions from horizontal federated learning. In SL, each client holds only a subset of features for the same samples, while labels are typically owned by a server or trusted party—for example, different hospitals may separately maintain medical imaging, laboratory test results, or clinical histories for the same patients, yet seek to jointly train disease prediction or clinical decision support models \cite{li2024split,guo2024split}. In a typical SL workflow, clients train local bottom models to transform private features into intermediate embeddings that are sent to the server, which then trains a top model using the labels and returns gradients to update the bottom models, enabling end-to-end collaborative training without exposing raw features \cite{thapa2022splitfed,vepakomma2018split,singh2019detailed}.

While SL avoids explicit data sharing, it exposes a new attack surface at the embedding interface between clients and the server, making it particularly susceptible to poisoning attacks with backdoor objectives~\cite{bai2023villain,he2023backdoor,naseri2024badvfl}. By compromising participating clients, adversaries can manipulate the intermediate representations exchanged during training, implanting hidden patterns that bias the top model toward attacker-defined predictions under specific trigger conditions. Such attacks operate directly in the representation space and therefore remain difficult to detect using conventional data-level defenses; for example, VILLAIN~\cite{bai2023villain} exploits inferred label information to selectively corrupt embeddings in a highly stealthy manner, which cause difficult to detect the malicious embedding and provides limited robustness. Although recent approaches \cite{dou2026securesplit,cho2024vflip} such as VFLIP~\cite{cho2024vflip} introduce SL-specific identification and purification mechanisms, they are still insufficient to reliably defend against adaptive backdoor attacks.

\begin{figure*}[ht]
  \centering

  \begin{minipage}[b]{0.23\linewidth}
    \centering
    \includegraphics[width=\linewidth]{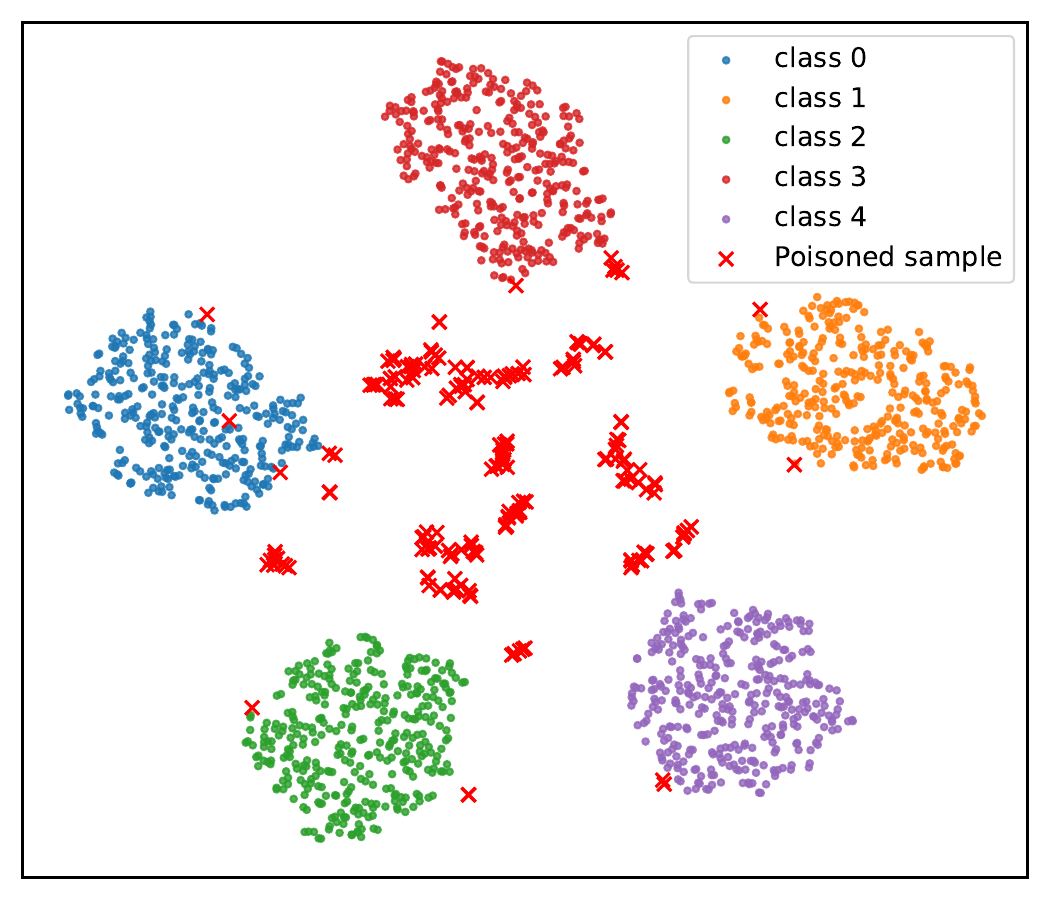}
    {\small (a) Embedding distribution w/ ViLLAIN attack}
  \end{minipage}
  \hfill
  \begin{minipage}[b]{0.23\linewidth}
    \centering
    \includegraphics[width=\linewidth]{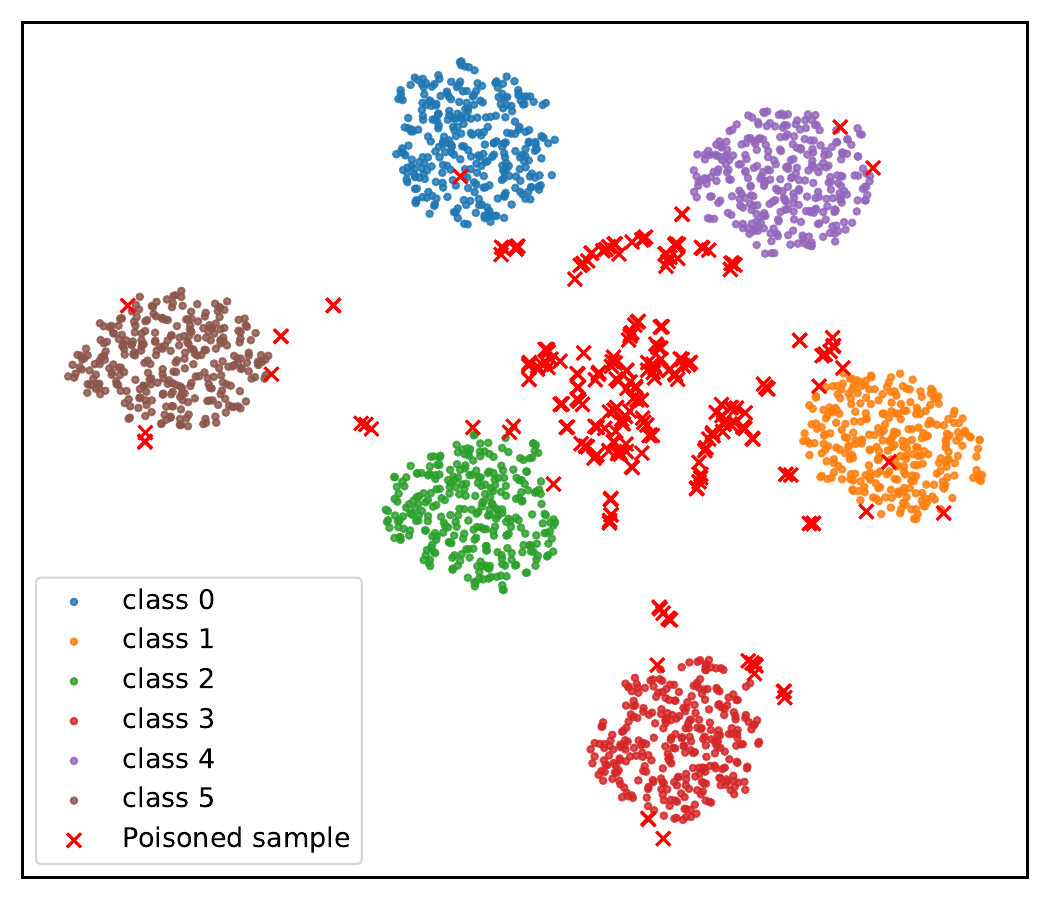}
    {\small (b) Embedding distribution w/ SplitNN attack}
  \end{minipage}
  \hfill
  \begin{minipage}[b]{0.23\linewidth}
    \centering
    \includegraphics[width=\linewidth]{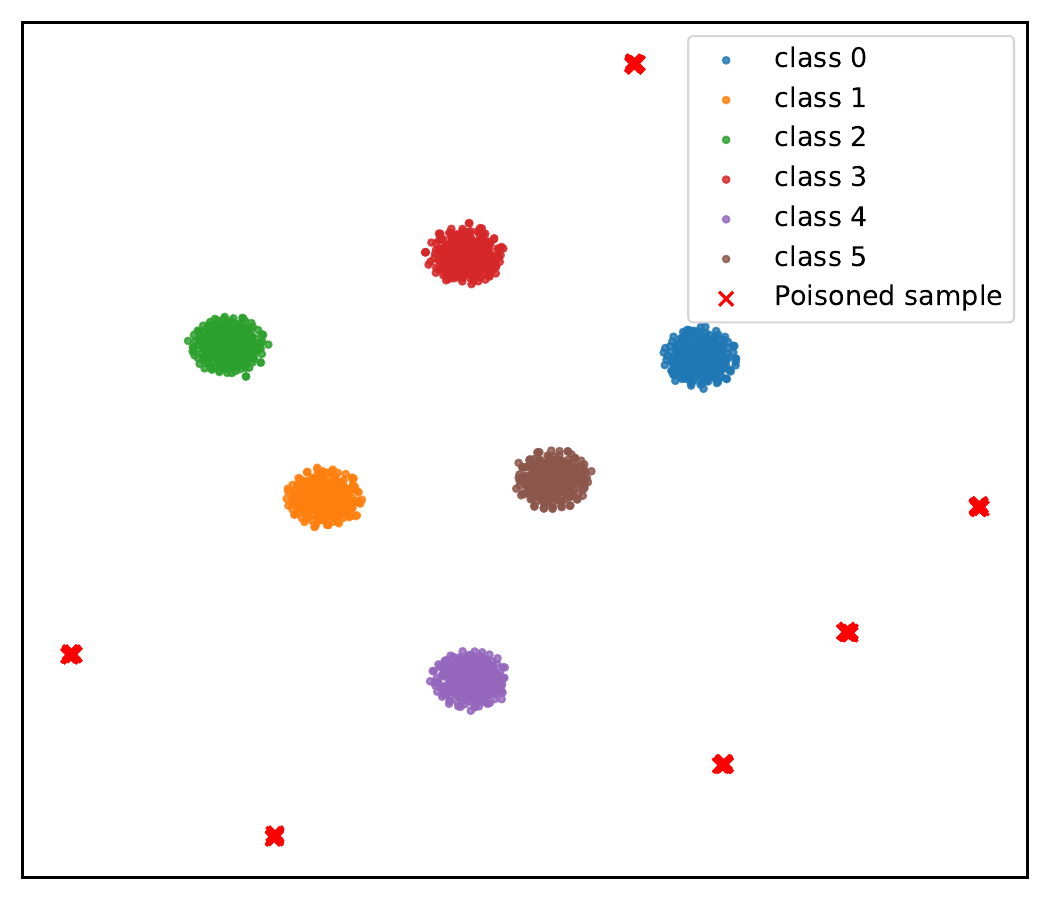}
    {\small (c) Consistency representation distribution w/ ViLLAIN attack}
  \end{minipage}
  \hfill
  \begin{minipage}[b]{0.23\linewidth}
    \centering
    \includegraphics[width=\linewidth]{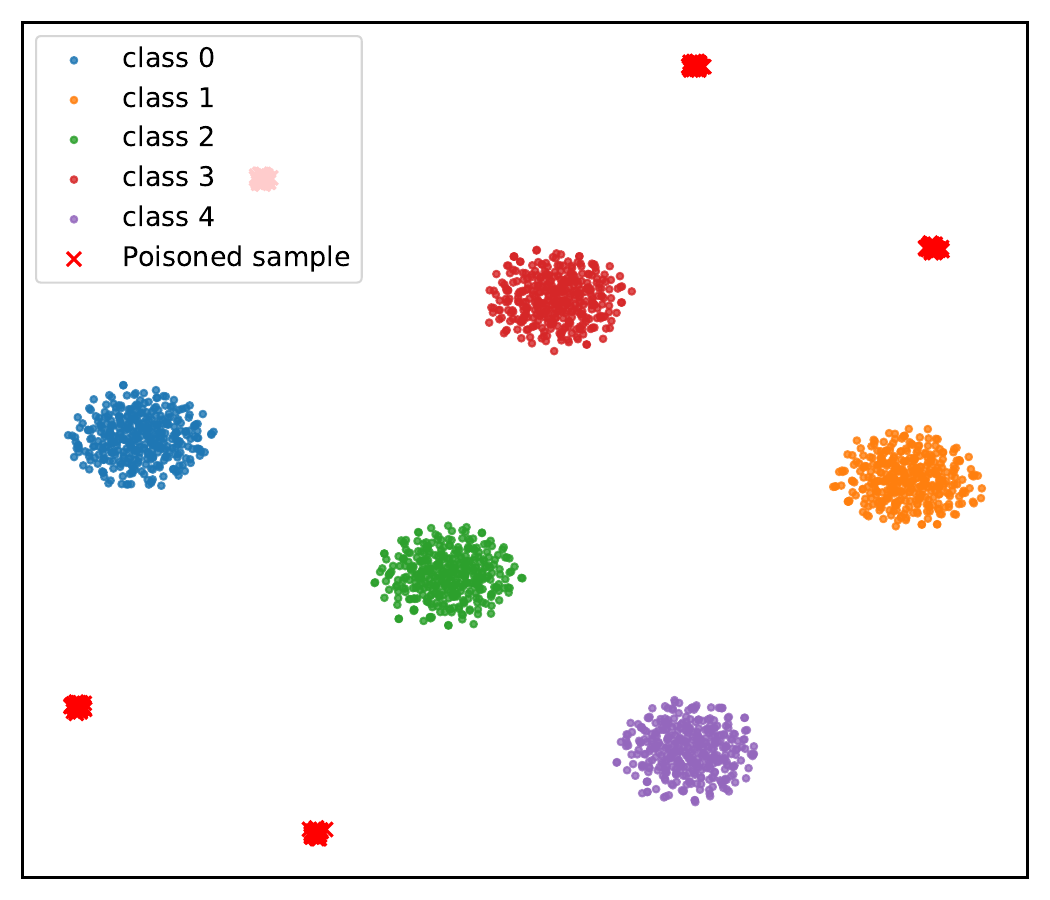}
    {\small (d) Consistency representation distribution w/ SplitNN attack}
  \end{minipage}

  \caption{Visualization of embedding distributions under backdoor attacks before and after prototype-consistency transformation.
(a) and (b) show the embedding distributions in the original embedding space under the ViLLAIN and SplitNN attacks, respectively, where poisoned embeddings are highly overlapped with benign ones and thus difficult to distinguish.
(c) and (d) illustrate the embedding distributions in the prototype-consistency representation space under the same attacks. After transforming embeddings based on their semantic consistency with class prototypes, poisoned samples become clearly separable from benign samples, highlighting the effectiveness of the proposed representation.}
  \label{fig:effect}
\end{figure*}

These backdoors are often highly stealthy \cite{bai2023villain,he2023backdoor}, causing the poisoned samples to have data distributions that highly overlap with those of clean samples, as shown in Fig \ref{fig:effect}(a) and (b).
To improve the robustness of split learning against stealthy backdoor attacks, we propose \alg, a
server-side defense mechanism that detects and removes poisoned embeddings by explicitly
exploiting \emph{class-conditional representation consistency} in the embedding space.
Our key observation in Fig \ref{fig:effect}(a) and (b) is that, while backdoor attacks can be carefully designed to evade
conventional anomaly detection, poisoned embeddings inevitably disrupt the semantic consistency
shared by benign samples within the same class.
Therefore, we can achieve effective separation by modeling the semantic consistency of benign samples within the same class and treating samples that significantly deviate from this consistency structure as potential poisoned samples.
\alg follows a two-stage design.
First, leveraging the labels naturally available in split learning, the server constructs robust
class prototypes and transforms each embedding into a
\emph{prototype-consistency representation} that characterizes its relative semantic alignment
with all classes, rather than relying on absolute geometric properties in the original embedding
space.
During this process, benign samples can often be distinguished from poisoned samples, as shown in Figure \ref{fig:effect}(c) and (d).
Second, \alg applies a class-conditional and distribution-free conformal filtering strategy to
identify embeddings that significantly deviate from typical class-consistent patterns, enabling
robust detection without assuming any parametric form of embedding distributions. We conducted extensive experiments on three different datasets, CIFAR-10, SVHN, and Bank Marketing, and achieved state-of-the-art performance.

We summarize our main contributions as follows:
\begin{itemize}
    \item We identify class-conditional representation consistency as a fundamental property of
    benign embeddings in split learning and show how backdoor attacks disrupt this structure.
    \item We propose a prototype-based consistency representation that maps embeddings into a
    relational space, substantially enhancing the separability between benign and poisoned
    samples.
    \item We design a class-conditional, distribution-free filtering mechanism that effectively
    mitigates backdoor attacks while preserving benign embeddings and overall model performance.
\end{itemize}

\section{Related work}
\label{sec:rela}

\subsection{Background on Split Learning}

SL ~\cite{fu2022blindfl,romanini2021pyvertical,liu2020asymmetrical} is a collaborative learning framework designed for vertically partitioned data, where multiple clients jointly train a model without sharing raw features or labels. In this setting, each client holds only a subset of features for all training samples, while the labels are exclusively maintained by a central server. By decoupling feature ownership and label access, SL enables privacy-preserving collaboration among parties with heterogeneous data sources.

Formally, consider an SL system with \( n \) clients \( \{C_i\}_{i=1}^n \). For a training example \( k \), the complete feature vector \( \bm{x}_k \) is distributed across clients such that
$\bm{x}_k = \bigcup_{i=1}^{n} \hat{\bm{x}}_k^i$,
where \( \hat{\bm{x}}_k^i \) denotes the feature subset owned by client \( C_i \). During each training round, client \( C_i \) applies a local bottom model \( \mathcal{L}_i \) to generate an embedding
$\bm{E}_k^i = \mathcal{L}_i(\hat{\bm{x}}_k^i)$,
which is transmitted to the server. The server aggregates the received embeddings using an aggregation function \( \mathcal{A}(\cdot) \),
$\bm{E}_k = \mathcal{A}(\bm{E}_k^1, \ldots, \bm{E}_k^n)$,
and trains a top model in a supervised manner with the corresponding label \( y_k \). Gradients with respect to each client’s embedding are then computed and sent back to update the bottom models. This iterative process continues until convergence and naturally extends to mini-batch training in practice.

\subsection{Backdoor Attacks against Split Learning}

SL enables multiple clients to collaboratively train a model by exchanging intermediate embeddings with a central server rather than sharing raw data. Despite its privacy advantages, this decentralized and communication-intensive training paradigm introduces significant security risks, making SL vulnerable to backdoor attacks~\cite{he2023backdoor,bai2023villain,naseri2024badvfl}. In such attacks, adversaries aim to implant hidden behaviors into the server-side model, causing it to output attacker-specified predictions when particular trigger patterns are present in the input.

Backdoor attacks in split learning (SL) are typically launched by malicious clients via data-level or representation-level manipulation. Data-level attacks poison local datasets with imperceptible triggers to bias training, as in BadVFL~\cite{naseri2024badvfl}. More stealthy representation-level attacks directly manipulate embeddings sent to the server without altering raw inputs; VILLAIN~\cite{bai2023villain} exemplifies this approach by carefully tuning malicious embeddings to evade detection. Related work on SplitNN~\cite{he2023backdoor} further shows that a client can inject backdoors by poisoning embeddings without modifying labels. Collectively, these studies highlight the inherent vulnerability of SL under partial trust assumptions~\cite{he2023backdoor,bai2023villain,naseri2024badvfl}.

\section{Our \alg}
\label{sec:alg}
\subsection{Empirical Analysis and Motivation}

As illustrated in Fig. 1(a) and Fig. 1(b), under embedding-space backdoor attacks, benign and poisoned samples are highly overlapped in the original embedding space, making them difficult to distinguish using conventional geometric criteria. Although poisoned embeddings exhibit a subtle tendency to deviate from benign ones within the same class, this deviation is insufficient to form a clear separation. This observation suggests that backdoor attacks introduce latent semantic perturbations that disrupt intra-class consistency rather than causing large geometric shifts.

Motivated by this phenomenon, our \alg enhances the robustness of split learning (SL) against backdoor attacks by explicitly exploiting class-conditional representation consistency in the embedding space. Specifically, \alg operates in two steps. First, the server constructs robust class prototypes using the labels naturally available in SL and represents each embedding by its relative consistency with all class prototypes, thereby transforming embeddings into a relational, class-aware representation. Second, a class-conditional, distribution-free filtering mechanism is applied to identify and remove embeddings that significantly deviate from typical class-consistent patterns. As a result, poisoned embeddings are pushed away from benign ones, as shown in Fig. 1(c) and Fig. 1(d), enabling effective backdoor mitigation while preserving benign embeddings and overall model performance.

\subsection{Prototype-based Consistency Representation}

We consider a training set composed of $m$ samples, each of which is represented on the server
side by an aggregated feature embedding.
Let $
\bm{E} = \{ \bm{E}_1, \bm{E}_2, \ldots, \bm{E}_m \} $
denote the collection of all embeddings, where $\bm{E}_k \in \mathbb{R}^d$ is produced through
the embedding aggregation process.
In split learning, although raw features and local models remain private to the clients, the
server retains access to the ground-truth labels $y_k$ associated with each training example and
receives the corresponding embeddings during training.

Under backdoor attacks, the embedding collection observed at the server can include both
legitimate and maliciously modified representations.
To maintain stealthiness, poisoned embeddings are typically designed to bypass conventional
anomaly detection mechanisms~\cite{bai2023villain,he2023backdoor}, making them difficult to
distinguish using simple similarity- or norm-based criteria.

\paragraph{Class Prototype Construction}
To capture the semantic structure of each class, the server constructs a robust prototype for
every class.
Let $\mathcal{C}$ denote the set of classes.
For each class $c \in \mathcal{C}$, we define its prototype as the coordinate-wise median of
embeddings belonging to that class:
\begin{align}
\label{eq:prototype}
\bm{p}_c = \text{Median}\big( \{ \bm{E}_k \mid y_k = c \} \big),
\end{align}
where the median operator is applied independently to each coordinate.
Using the median ensures robustness against a small fraction of poisoned embeddings within
each class.

\begin{table*}[!t]
\caption{Comparison of Different Defense Methods Evaluated by MA and ASR, where larger MA and lower ASR indicate superior performance.}
\centering
\footnotesize
\begin{tabular}{cc|cccccccccccc}
\midrule[1.2pt]
\multirow{2}{*}{Dataset}        & \multirow{2}{*}{Attack} & \multicolumn{2}{c}{No defense} & \multicolumn{2}{c}{DP} & \multicolumn{2}{c}{MP} & \multicolumn{2}{c}{ANP} & \multicolumn{2}{c}{VFLIP} & \multicolumn{2}{c}{\alg} \\ \cline{3-14} 
                                &                         & MA             & ASR           & MA         & ASR       & MA         & ASR       & MA         & ASR        & MA          & ASR         & MA                 & ASR                \\ \hline
\multirow{3}{*}{CIFAR-10}       & VILLAIN attack          & 0.80           & 0.92          & 0.74       & 0.77      & 0.74       & 0.67      & 0.77       & 0.66       & 0.65        & 0.35        & \textbf{0.83}      & \textbf{0.06}      \\
                                & SplitNN attack          & 0.77           & 0.95          & 0.79       & 0.73      & 0.72       & 0.70      & 0.74       & 0.67       & 0.72        & 0.33        & \textbf{0.85}      & \textbf{0.05}      \\
                                & BadVFL                  & 0.79           & 0.89          & 0.77       & 0.68      & 0.72       & 0.78      & 0.76       & 0.65       & 0.66        & 0.28        & \textbf{0.84}      & \textbf{0.03}      \\ \hline
\multirow{3}{*}{SVHN}           & VILLAIN attack          & 0.83           & 0.95          & 0.80       & 0.91      & 0.82       & 0.88      & 0.80       & 0.75       & 0.76        & 0.29        & \textbf{0.87}      & \textbf{0.05}      \\
                                & SplitNN attack          & 0.80           & 0.95          & 0.78       & 0.93      & 0.74       & 0.82      & 0.77       & 0.69       & 0.75        & 0.33        & \textbf{0.85}      & \textbf{0.07}      \\
                                & BadVFL                  & 0.78           & 0.94          & 0.78       & 0.84      & 0.80       & 0.77      & 0.74       & 0.72       & 0.69        & 0.27        & \textbf{0.85}      & \textbf{0.04}      \\ \hline
\multirow{3}{*}{Bank marketing} & VILLAIN attack          & 0.84           & 0.95          & 0.82       & 0.89      & 0.75       & 0.52      & 0.77       & 0.67       & 0.73        & 0.39        & \textbf{0.85}      & \textbf{0.08}      \\
                                & SplitNN attack          & 0.82           & 0.93          & 0.79       & 0.84      & 0.77       & 0.49      & 0.75       & 0.62       & 0.69        & 0.35        & \textbf{0.82}      & \textbf{0.05}      \\
                                & BadVFL                  & 0.84           & 0.90          & 0.78       & 0.84      & 0.76       & 0.50      & 0.77       & 0.58       & 0.72        & 0.37        & \textbf{0.86}      & \textbf{0.07}      \\ \midrule[1.2pt]
\end{tabular}
\label{tab:main}
\end{table*}

\paragraph{Prototype Consistency Transformation}
Instead of directly analyzing embeddings in the original space, \alg transforms each embedding
into a \emph{prototype consistency representation} that characterizes its semantic alignment
with all class prototypes.
For each embedding $\bm{E}_k$, we construct a consistency vector:
\begin{align}
\label{eq:consistency_vector}
\bm{v}_k =
\Big[
\cos(\bm{E}_k, \bm{p}_1),
\cos(\bm{E}_k, \bm{p}_2),
\ldots,
\cos(\bm{E}_k, \bm{p}_{|\mathcal{C}|})
\Big],
\end{align}
where $\cos(\cdot,\cdot)$ denotes cosine similarity.

This transformation maps embeddings from the original feature space into a relational space
that encodes their relative similarities to all class prototypes.
For benign embeddings, the resulting vectors exhibit stable and class-consistent patterns.
In contrast, poisoned embeddings tend to produce anomalous similarity profiles due to the
semantic distortion introduced by backdoor triggers. A clear separation between benign and poisoned samples can be observed.

\paragraph{Consistency Deviation Score}
For each class $c \in \mathcal{C}$, we further compute a class-specific reference pattern:
\begin{align}
\bm{\mu}_c = \text{Median}\{ \bm{v}_k \mid y_k = c \}.
\end{align}
The nonconformity (consistency deviation) score of an embedding $\bm{E}_k$ is then defined as:
\begin{align}
s_k = \left\| \bm{v}_k - \bm{\mu}_{y_k} \right\|_2.
\end{align}
A larger $s_k$ indicates that $\bm{E}_k$ deviates more from the typical relational behavior of
its class and is therefore more likely to be poisoned.

\paragraph{Conformal Filtering}
To determine whether an embedding is benign, we adopt a conformal filtering strategy that
evaluates each embedding based on its relative rank among samples of the same class.
This rank-based formulation does not assume any parametric distribution of the deviation scores
and only relies on their ordering within each class, making it robust to scale variations and
class imbalance.

For class $c$, let
$\mathcal{S}_c = \{ s_k \mid y_k = c \}$
denote the set of consistency deviation scores.
For an embedding $\bm{E}_k$ with label $y_k = c$, we compute its conformal $p$-value as:
\begin{align}
p_k = \frac{|\{ s \in \mathcal{S}_c : s \ge s_k \}| + 1}{|\mathcal{S}_c| + 1}.
\end{align}

Given a predefined significance level $\alpha \in (0,1)$, an embedding is classified as benign if $p_k > \alpha$.
Under certain constraints, we can guarantee the security of \alg in the vertical split model. Specifically, as long as the required assumptions hold, \alg does not leak sensitive information of the participating parties during model training and inference, thereby ensuring the overall system security.

\begin{algorithm}[t]
    \caption{\alg.}
    \label{algo:ouralg}
    \begin{algorithmic}[1]
        \renewcommand{\algorithmicrequire}{\textbf{Input:}}
        \renewcommand{\algorithmicensure}{\textbf{Output:}}
        \Require Full embedding set $\bm{E}=\{\bm{E}_k\}_{k=1}^m$, label set $\bm{y}$, significance level $\alpha$.
        \Ensure Benign embedding index set $\mathcal{B}$.
        
        \State $\mathcal{B} \leftarrow \emptyset$.
        
        \State \textcolor{blue}{// Step I: Class Prototype Construction}
        \For{each class $c \in \mathcal{C}$}
            \State $\mathcal{I}_c \leftarrow \{ k \mid y_k = c \}$.
            \State Compute class prototype $\bm{p}_c \gets \text{Median}(\{\bm{E}_k \mid k \in \mathcal{I}_c\})$.
        \EndFor
        
        \State \textcolor{blue}{// Step II: Prototype Consistency Representation}
        \For{each embedding $\bm{E}_k \in \bm{E}$}
            \State Construct consistency vector
            $\bm{v}_k \gets [\cos(\bm{E}_k,\bm{p}_1), \ldots, \cos(\bm{E}_k,\bm{p}_{|\mathcal{C}|})]$.
        \EndFor
        
        \State \textcolor{blue}{// Step III: Class-Conditional Deviation Scoring}
        \For{each class $c \in \mathcal{C}$}
            \State Compute reference pattern $\bm{\mu}_c \gets \text{Median}(\{\bm{v}_k \mid k \in \mathcal{I}_c\})$.
        \EndFor
        \For{each embedding $\bm{E}_k \in \bm{E}$}
            \State Compute deviation score $s_k \gets \|\bm{v}_k - \bm{\mu}_{y_k}\|_2$.
        \EndFor
        
        \State \textcolor{blue}{// Step IV: Class-Conditional Conformal Filtering}
        \For{each class $c \in \mathcal{C}$}
            \State $\mathcal{S}_c \gets \{ s_k \mid k \in \mathcal{I}_c \}$.
            \For{each $k \in \mathcal{I}_c$}
                \State Compute conformal $p$-value
                $p_k \gets \dfrac{|\{ s \in \mathcal{S}_c : s \ge s_k \}| + 1}{|\mathcal{S}_c| + 1}$.
                \If{$p_k > \alpha$}
                    \State $\mathcal{B} \leftarrow \mathcal{B} \cup \{k\}$.
                \EndIf
            \EndFor
        \EndFor
        
    \end{algorithmic}
\end{algorithm}

\section{Experiment}
\label{sec:exp}

\subsection{Experiment setting}

We conduct comprehensive experimental evaluations of \alg on four widely used image classification benchmarks: CIFAR-10, SVHN and Bank Market \cite{moro2014data}. These datasets span diverse characteristics and data modalities (including text and images), varying class complexity, and different data scales, enabling a comprehensive evaluation of the robustness and generalization ability of \alg under diverse learning scenarios.
To rigorously assess the robustness of \alg against malicious behaviors, we evaluate its performance under several backdoor poisoning attacks including VILLAIN~\cite{bai2023villain}, BadVFL~\cite{naseri2024badvfl}, and the SplitNN backdoor attack~\cite{he2023backdoor}.
We employ a split model architecture in which the client-side local model is a 4-layer fully connected network (FCN) for the CIFAR-10, SVHN, and Bank Marketing datasets, while the server-side model is consistently implemented as a 3-layer FCN across all datasets. The architecture of neural network can seen in supplement file. All models are trained for 80 communication rounds, with the poisoning process introduced at the 20th round to ensure a consistent adversarial setting. The learning rate is fixed at $1 \times 10^{-2}$, and a uniform batch size of 5000 is used across all experiments to reduce the impact of hyperparameter variations. Additionally, the trigger magnitude is set to 1.0 to balance attack effectiveness and training stability, enabling a controlled and reproducible evaluation of model robustness under poisoning attacks. Our default $\alpha$ is 0.5.
We defaulted to using four local clients, one of which is maclious. The poison rate was set to 5\%.
In addition, we compare \alg with four common defense baselines in split learning: differential privacy (DP)~\cite{abadi2016deep}, model pruning (MP)~\cite{liu2018fine}, adversarial neuron pruning (ANP)~\cite{wu2021adversarial}, and VFLIP~\cite{cho2024vflip}, a recent method for vertical federated learning. Safesplit~\cite{rieger2025safesplit} targets U-shaped SL and is not included due to incompatible settings.

We adopt two standard evaluation metrics to measure both model utility and attack effectiveness: main accuracy (MA) and attack success rate (ASR). MA is defined as the proportion of clean test samples that are correctly classified by the trained model, reflecting its predictive performance on benign data. ASR measures the fraction of trigger-injected test samples that are misclassified into the attacker-specified target class, characterizing the strength of the backdoor. An effective defense should maintain high ACC while keeping ASR low, ensuring robustness without sacrificing accuracy.

\subsection{Main results}


Table \ref{tab:main} presents a comprehensive comparison of different defense methods across three datasets (CIFAR-10, SVHN, and Bank Marketing) under multiple attack scenarios, where a higher MA and a lower ASR indicate better performance. Without any defense, the ASR remains extremely high (close to or above 0.9) across all datasets and attack types, demonstrating the severe vulnerability of the models. Existing defenses, including DP, MP, ANP, and VFLIP, can partially reduce ASR, but often at the cost of noticeable performance degradation in MA. In particular, DP and MP suffer from an evident utility–privacy trade-off, while ANP provides limited robustness improvements. Although VFLIP achieves lower ASR in some cases, it incurs a significant drop in MA, especially on CIFAR-10 and SVHN. In contrast, our method \alg consistently achieves the best or near-best MA while dramatically reducing ASR across all datasets and attack settings. Specifically, \alg reduces ASR to as low as 0.03–0.08 while maintaining or even improving MA compared to the no-defense baseline. These results demonstrate that \alg effectively balances attack mitigation and model utility, providing robust and stable protection against diverse attacks in vertical federated learning.

\subsection{Ablation analysis}

\begin{table}[!t]
\tiny
\caption{Ablation Study of \alg.}
\centering
\begin{tabular}{l|cc|cc|cc}
\midrule[1.2pt]
\multirow{2}{*}{Method} 
& \multicolumn{2}{c}{VILLAIN} 
& \multicolumn{2}{c}{SplitNN} 
& \multicolumn{2}{c}{BadVFL} \\ \cline{2-7}
& MA & ASR & MA & ASR & MA & ASR \\ \hline
\alg
& \textbf{0.83} & \textbf{0.06} 
& \textbf{0.85} & \textbf{0.05} 
& \textbf{0.84} & \textbf{0.03} \\

\alg$_\text{w/o Consistency Rep}.$ 
& 0.79 & 0.41 
& 0.81 & 0.48 
& 0.80 & 0.35 \\

\alg$_\text{w/o All-class Relation}.$
& 0.80 & 0.36 
& 0.82 & 0.34 
& 0.81 & 0.21 \\

\alg$_\text{w/o Class-conditional}.$
& 0.81 & 0.23 
& 0.83 & 0.27 
& 0.82 & 0.19 \\
\midrule[1.2pt]
\end{tabular}
\label{tab:ablation_cifar10}
\end{table}

Table~\ref{tab:ablation_cifar10} reports the ablation results of \alg on CIFAR-10 under three representative backdoor attack scenarios: VILLAIN, SplitNN, and BadVFL. The complete \alg consistently achieves the best performance, yielding the highest model accuracy (MA) and the lowest attack success rate (ASR) across all settings, which verifies the overall effectiveness of our design. Removing any key component leads to performance degradation, indicating that these modules are complementary. In particular, excluding the consistency representation results in the most severe increase in ASR, highlighting its critical role in learning robust and attack-invariant features. Eliminating the all-class relation or the class-conditional modeling moderately reduces MA and noticeably increases ASR, suggesting that both global inter-class relationships and fine-grained class-specific constraints contribute to enhanced robustness. 

\subsection{Impact of various neural network architectures on bottom models.}

\begin{table}[h]
\centering
\footnotesize
\caption{An Analysis of the Impact of Different Bottom Models under the VILLAIN Attack Using the CIFAR-10 Dataset.}
\label{tab:cifar10_backbone}
\setlength{\tabcolsep}{5pt}
\renewcommand{\arraystretch}{1.1}
\begin{tabular}{lccccc}
\toprule
\multirow{2}{*}{Method} & \multicolumn{2}{c}{ResNet-18} & \phantom{a} & \multicolumn{2}{c}{VGG-19} \\
\cmidrule{2-3} \cmidrule{5-6}
 & MA & ASR && MA & ASR \\
\midrule
No Defense     &  0.86 &  0.90 &&  0.83 &  0.87 \\
DP             &  0.82 &  0.77 &&  0.84 &  0.78 \\
MP             &  0.79 &  0.62 &&  0.77 &  0.63 \\
ANP            &  0.82 &  0.64 &&  0.77 &  0.59\\
VFLIP          &  0.74 &  0.37 &&  0.68 &  0.27 \\
\midrule
\alg  & \textbf{0.88} & \textbf{0.05} && \textbf{0.87} & \textbf{0.04} \\
\bottomrule
\end{tabular}
\label{tab:various_nn}
\end{table}

Table \ref{tab:various_nn} shows that under the VILLAIN attack on CIFAR-10, ResNet-18 and VGG-19 exhibit consistent trends across defenses, though with different magnitudes. Without defense, both models achieve high MA and ASR, indicating severe vulnerability. DP, MP, and ANP reduce ASR only marginally and at the cost of degraded MA, providing limited protection in VFL. VFLIP substantially lowers ASR but incurs a notable accuracy drop, revealing a security–utility trade-off. In contrast, the proposed \alg achieves the highest MA and lowest ASR under both architectures, demonstrating superior effectiveness and robust generalization across bottom models.

\subsection{Sensitive analysis}

Due to space constraints, \textbf{we place the sensitivity analysis of the poisoning rate, malicious client rate, and the initial poisoning round in the supplement file}.

\myparatight{Sensitive analysis of filtering parameter $\alpha$} As shown in Table~\ref{tab:sens}, the performance of the system is sensitive to the choice of $\alpha$. 
When $\alpha = 0.3$, all three attacks achieve relatively high ASR, indicating that the defense is ineffective under a small $\alpha$. 
Increasing $\alpha$ to 0.5 leads to a sharp reduction in ASR across VILLAIN, SplitNN attack, and BadVFL, while the model accuracy (MA) reaches its highest or near-highest level, suggesting an optimal trade-off between security and utility. 
When $\alpha$ increases further to 0.7, the defense remains effective with consistently low ASR, but MA shows a slight degradation. 
In contrast, setting $\alpha = 0.9$ causes a significant drop in MA without bringing additional security benefits, as ASR remains low. 
Overall, these results indicate that a moderate value of $\alpha$, particularly $\alpha = 0.5$, provides the best balance between model performance and robustness against different attacks.

\begin{table}[h]
\centering
\footnotesize
\caption{Sensitive analysis of $\alpha$.}
\begin{tabular}{c|cccccc}
\midrule[1.2pt]
\multirow{2}{*}{$\alpha$} & \multicolumn{2}{c}{VILLAIN attack} & \multicolumn{2}{c}{SplitNN attack} & \multicolumn{2}{c}{BadVFL} \\ \cline{2-7} 
                          & MA               & ASR             & MA               & ASR             & MA           & ASR         \\ \hline
0.3                       & 0.74             & 0.68            & 0.74             & 0.69            & 0.82         & 0.46        \\
0.5                       & 0.83             & 0.06            & 0.85             & 0.05            & 0.84         & 0.03        \\
0.7                       & 0.82             & 0.05            & 0.82             & 0.07            & 0.83         & 0.04        \\
0.9                       & 0.57             & 0.09            & 0.52             & 0.04            & 0.52         & 0.03        \\ \midrule[1.2pt]
\end{tabular}
\label{tab:sens}
\end{table}

\begin{figure*}
    \centering
    \includegraphics[width=0.8\linewidth]{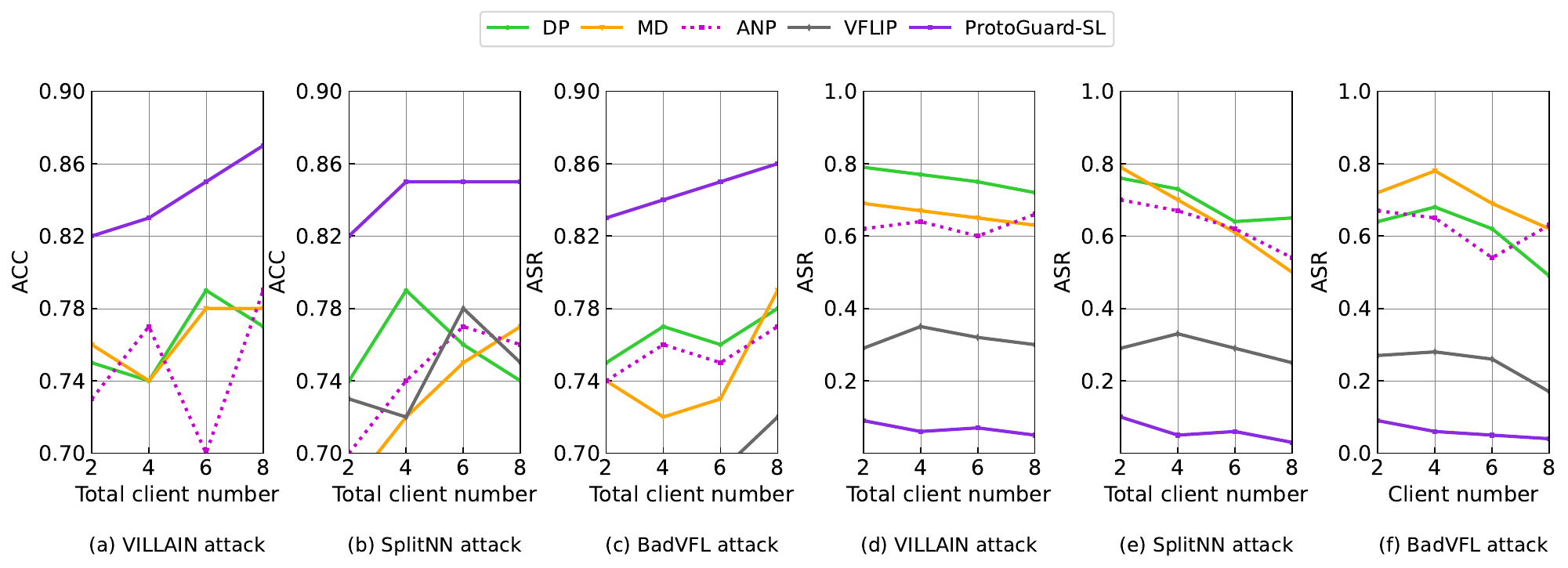} 
    \caption{Impact of the total client number, where CIFAR-10 dataset is considered.}
    \label{total_num}
           \vspace{-0.25cm}
\end{figure*}

\myparatight{Impact of total client number}We first investigate the impact of the total number of participating clients on the robustness of ProtoGuard-SL. As shown in Fig.\ref{total_num}, we vary the total client number from 2 to 8 under different attack settings while keeping the proportion of malicious clients fixed. The results demonstrate that ProtoGuard-SL consistently maintains high model MA and low ASR as the number of clients increases. In contrast, baseline defenses exhibit noticeable performance degradation, especially in terms of ASR, when more clients participate in the training process. This is because a larger number of clients introduces more heterogeneous embeddings, which can amplify the effect of poisoned representations for conventional defenses. Benefiting from its class-conditional prototype consistency modeling, ProtoGuard-SL effectively suppresses poisoned embeddings even in large-scale client settings, indicating good scalability and robustness with respect to the total client number.

\section{Conclusion}

In this paper, we study embedding-space backdoor attacks in vertical split learning and show that poisoned embeddings violate class-conditional representation consistency. Leveraging this insight, we propose \alg, a server-side defense that maps embeddings to a prototype-consistent space and applies a class-conditional, distribution-free filter to detect and remove anomalies. Extensive experiments across datasets and attack settings demonstrate that \alg significantly reduces attack success while maintaining benign accuracy, consistently outperforming existing defenses.

\section{Acknowledgements}
The work was partially supported by the following: 
The Zhejiang Provincial Natural Science Foundation - Exploration Project under No. LMS26F020007,
the Wenzhou Applied Fundamental Research Program (Basic Research) under No. GG20250198,
the WKU 2026 International Frontier Interdisciplinary Research Institute Talent Program under No. WKUTP2026002,
the WKU 2025 International Collaborative Research Program under No. ICRPSP2025001.

\bibliographystyle{IEEEbib}
\bibliography{icme2026references}

@inproceedings{rieger2025safesplit,
  title={SafeSplit: A Novel Defense Against Client-Side Backdoor Attacks in Split Learning},
  author={Rieger, Phillip and Pegoraro, Alessandro and Kumari, Kavita and Abera, Tigist and Knauer, Jonathan and Sadeghi, Ahmad-Reza},
  booktitle={NDSS},
  year={2025}
}

@inproceedings{bai2023villain,
  title={VILLAIN: Backdoor attacks against vertical split learning},
  author={Bai, Yijie and Chen, Yanjiao and Zhang, Hanlei and Xu, Wenyuan and Weng, Haiqin and Goodman, Dou},
  booktitle={USENIX Security Symposium},
  year={2023}
}

@article{he2023backdoor,
  title={Backdoor attack against split neural network-based vertical federated learning},
  author={He, Ying and Shen, Zhili and Hua, Jingyu and Dong, Qixuan and Niu, Jiacheng and Tong, Wei and Huang, Xu and Li, Chen and Zhong, Sheng},
  journal={IEEE Transactions on Information Forensics and Security},
  volume={19},
  pages={748--763},
  year={2023},
  publisher={IEEE}
}

@inproceedings{mcmahan2017communication,
Author = {H. Brendan McMahan and Eider Moore and Daniel Ramage and Seth Hampson and Blaise Ag{\"u}era y Arcas},
Booktitle = {AISTATS},
Title = {Communication-Efficient Learning of Deep Networks from Decentralized Data},
Year = {2017}
}

@inproceedings{fu2022blindfl,
  title={Blindfl: Vertical federated machine learning without peeking into your data},
  author={Fu, Fangcheng and Xue, Huanran and Cheng, Yong and Tao, Yangyu and Cui, Bin},
  booktitle={SIGMOD},
  year={2022}
}

@article{romanini2021pyvertical,
  title={Pyvertical: A vertical federated learning framework for multi-headed splitnn},
  author={Romanini, Daniele and Hall, Adam James and Papadopoulos, Pavlos and Titcombe, Tom and Ismail, Abbas and Cebere, Tudor and Sandmann, Robert and Roehm, Robin and Hoeh, Michael A},
  journal={arXiv preprint arXiv:2104.00489},
  year={2021}
}

@article{liu2020asymmetrical,
  title={Asymmetrical vertical federated learning},
  author={Liu, Yang and Zhang, Xiong and Wang, Libin},
  journal={arXiv preprint arXiv:2004.07427},
  year={2020}
}

@article{naseri2024badvfl,
  title={Badvfl: Backdoor attacks in vertical federated learning},
  author={Naseri, Mohammad and Han, Yufei and De Cristofaro, Emiliano},
  booktitle={2024 IEEE Symposium on Security and Privacy (SP)},
  pages={2013--2028},
  year={2024},
  organization={IEEE}
}

@inproceedings{abadi2016deep,
  title={Deep learning with differential privacy},
  author={Abadi, Martin and Chu, Andy and Goodfellow, Ian and McMahan, H Brendan and Mironov, Ilya and Talwar, Kunal and Zhang, Li},
  booktitle={Proceedings of the 2016 ACM SIGSAC conference on computer and communications security},
  pages={308--318},
  year={2016}
}

@inproceedings{liu2018fine,
  title={Fine-pruning: Defending against backdooring attacks on deep neural networks},
  author={Liu, Kang and Dolan-Gavitt, Brendan and Garg, Siddharth},
  booktitle={International symposium on research in attacks, intrusions, and defenses},
  pages={273--294},
  year={2018},
  organization={Springer}
}

@article{wu2021adversarial,
  title={Adversarial neuron pruning purifies backdoored deep models},
  author={Wu, Dongxian and Wang, Yisen},
  journal={Advances in Neural Information Processing Systems},
  volume={34},
  pages={16913--16925},
  year={2021}
}

@inproceedings{cho2024vflip,
  title={VFLIP: A Backdoor Defense for Vertical Federated Learning via Identification and Purification},
  author={Cho, Yungi and Han, Woorim and Yu, Miseon and Lee, Younghan and Bae, Ho and Paek, Yunheung},
  booktitle={European Symposium on Research in Computer Security},
  year={2024}
}

@inproceedings{thapa2022splitfed,
  title={Splitfed: When federated learning meets split learning},
  author={Thapa, Chandra and Arachchige, Pathum Chamikara Mahawaga and Camtepe, Seyit and Sun, Lichao},
  booktitle={AAAI},
  year={2022}
}

@article{poirot2019split,
  title={Split learning for collaborative deep learning in healthcare},
  author={Poirot, Maarten G and Vepakomma, Praneeth and Chang, Ken and Kalpathy-Cramer, Jayashree and Gupta, Rajiv and Raskar, Ramesh},
  journal={arXiv preprint arXiv:1912.12115},
  year={2019}
}

@article{vepakomma2018split,
  title={Split learning for health: Distributed deep learning without sharing raw patient data},
  author={Vepakomma, Praneeth and Gupta, Otkrist and Swedish, Tristan and Raskar, Ramesh},
  journal={arXiv preprint arXiv:1812.00564},
  year={2018}
}

@article{singh2019detailed,
  title={Detailed comparison of communication efficiency of split learning and federated learning},
  author={Singh, Abhishek and Vepakomma, Praneeth and Gupta, Otkrist and Raskar, Ramesh},
  journal={arXiv preprint arXiv:1909.09145},
  year={2019}
}

@article{moro2014data,
  title={A data-driven approach to predict the success of bank telemarketing},
  author={Moro, S{\'e}rgio and Cortez, Paulo and Rita, Paulo},
  journal={Decision Support Systems},
  volume={62},
  pages={22--31},
  year={2014},
  publisher={Elsevier}
}

@inproceedings{li2024split,
  title={Split learning for distributed collaborative training of deep learning models in health informatics},
  author={Li, Zhuohang and Yan, Chao and Zhang, Xinmeng and Gharibi, Gharib and Yin, Zhijun and Jiang, Xiaoqian and Malin, Bradley A},
  booktitle={AMIA Annual Symposium Proceedings},
  volume={2023},
  pages={1047},
  year={2024}
}

@inproceedings{guo2024split,
  title={Split Learning Optimized For The Medical Field: Reducing Communication Overhead},
  author={Guo, Shuai and Lu, Zhi and Lu, Songfeng and Cui, Yongquan and Tang, Xueming and Wu, Junjun},
  booktitle={2024 IEEE International Conference on Bioinformatics and Biomedicine (BIBM)},
  pages={3226--3231},
  year={2024},
  organization={IEEE}
}

@article{dou2026securesplit,
  title={SecureSplit: Mitigating Backdoor Attacks in Split Learning},
  author={Dou, Zhihao and Cui, Dongfei and Wang, Weida and Gao, Anjun and Quan, Yueyang and Ma, Mengyao and Vo, Viet and Bai, Guangdong and Liu, Zhuqing and Fang, Minghong},
  journal={WWW},
  year={2026}
}

\end{document}